\documentclass[aps,prd,12pt,nofootinbib,tightenlines,floatfix,notitlepage]{revtex4-1}
\usepackage[utf8]{inputenc}

\usepackage{amsmath}
\usepackage{amssymb}
\usepackage{graphicx}

\newcommand{\GeV}{\ensuremath\text{ GeV}}
\newcommand{\di}{\ensuremath\mathrm{d}}

\begin{document}
\preprint{KA-TP-08-2017}
\title{Jet clustering dependence of VBF Higgs production}
\author{Michael Rauch}
\affiliation{Institute for Theoretical Physics, Karlsruhe Institute of
Technology (KIT), Germany}
\author{Dieter Zeppenfeld}
\affiliation{Institute for Theoretical Physics, Karlsruhe Institute of
Technology (KIT), Germany}

\begin{abstract}
Precise predictions for Higgs production via vector-boson fusion play an
important role when testing the properties of the Higgs boson and
probing new-physics effects. While the inclusive cross section changes 
little when including  NNLO and N3LO QCD corrections, a differential NNLO
calculation with typical VBF cuts~\cite{Cacciari:2015jma} shows large 
corrections of up to 10\% in distributions.

In this article, we investigate the dependence of the differential NNLO QCD calculation
on the jet definition. Starting from the known results at a fixed jet
clustering choice, we use the electroweak $H$+3~jets production cross 
section at NLO precision to derive NNLO results for $H$+2~jets production 
for other jet definitions. We find that larger clustering radii
significantly reduce the impact of NNLO corrections. The sizable NNLO
corrections for distributions are largely caused by the broader energy flow
inside jets at NNLO and are to be expected generally for processes with jets
at LO.
\end{abstract}

\maketitle

\section{Introduction}

With the discovery of the Higgs
boson~\cite{Aad:2012tfa,Chatrchyan:2012xdj} at the Large Hadron Collider
(LHC) at CERN in 2012, all particles predicted by the Standard Model
(SM) have finally been observed. One of the remaining tasks is to verify
that it is indeed the SM Higgs which has been observed. For this, a
detailed study of its properties, in particular its couplings, needs to
be carried out with high precision. 

The vector-boson-fusion (VBF) production mode of the Higgs boson is a
crucial ingredient for this
task~\cite{Cahn:1983ip,Dawson:1984gx,Duncan:1985vj,Cahn:1986zv,Kleiss:1987cj,Barger:1988mr,Butterworth:2002tt,Dittmaier:2011ti,Dittmaier:2012vm,Heinemeyer:2013tqa,Rauch:2016pai}.
Its cross section is sizable, being the second largest one for Higgs
production, after gluon fusion. The process exhibits a characteristic
feature, two energetic jets in the forward and backward regions of the 
detector, the
so-called tagging jets~\cite{Zeppenfeld:1999yd}. The decay products of
the Higgs boson are typically more central and in-between them. Also,
the Higgs boson is typically produced with considerable transverse
momentum. These can be exploited in the kinematic reconstruction of the
decay products, e.g.\ in decays into $\tau$
pairs~\cite{Plehn:1999xi,Aad:2015vsa,Chatrchyan:2014nva}, or for
invisible decay
modes~\cite{Eboli:2000ze,Aad:2015txa,Khachatryan:2016whc}.
Additionally, the nature of the $HVV$ vertex can be probed through the
azimuthal-angle distribution of the tagging
jets~\cite{Plehn:2001nj,Andersen:2010zx}.

In order to make full use of the LHC data, precise knowledge and good 
understanding of the SM prediction of VBF-$H$ production is crucial.
The next-to-leading order~(NLO) QCD 
corrections~\cite{Han:1992hr,Figy:2003nv,Berger:2004pca} have been known
for quite some time and yield corrections of $\mathcal{O}(10\%)$.
Choosing the momentum transfer of the exchanged virtual bosons as
central scale proves to be a good choice to minimize the QCD corrections. 
The NLO EW corrections are of a similar 
size~\cite{Ciccolini:2007jr,Ciccolini:2007ec,Figy:2010ct},
with opposite sign for inclusive cross sections.
Inclusive cross sections have been calculated to next-to-next-to-leading
order~(NNLO) QCD in Refs.~\cite{Bolzoni:2010xr,Bolzoni:2011cu} and to
next-to-next-to-next-to-leading order~(N3LO) in Ref.~\cite{Dreyer:2016oyx},
using the structure-function approach in both cases. The effects are
around the percent level for NNLO and at a few per mille for N3LO. The
latter further reduces the associated scale uncertainty. 

NNLO effects on differential distributions have been studied in the VBF
approximation~\cite{Cacciari:2015jma}, where QCD corrections to the two 
quark lines are considered as independent, analogous to the 
structure-function approach. The effects turn out to be
sizable, yielding a reduction of around 6\% for the fiducial cross section with 
typical VBF cuts (high invariant mass and large rapidity separation of the 
tagging jet pair), and even larger corrections are found for  
distributions which has been surprising, given the smallness of corrections 
for inclusive cross sections.

One should bear in mind, however, that the VBF cross sections at LO are
completely independent of the jet definition scheme, due to the large rapidity
separation enforced by the VBF cuts. Even in an NLO cross section, the 
internal jet shape
is only modeled by up to two partons, i.e.\ any dependence on the jet algorithm or the
jet radius in the rapidity-azimuthal angle-plane is modeled at LO only.
Indeed it was found long time ago that the jet shape of current jets 
in DIS, which have the same properties as the tagging jets in VBF, is much
wider at NLO, when up to three partons model the internal energy
flow~\cite{Kauer:1999px}. These effects will lead to differences in 
jet algorithm and jet radius dependencies for NNLO as compared to NLO cross
sections.

In this paper, we investigate the dependence of the differential NNLO
cross section on the definition of the jets. We test how different
choices change the size of the corrections, and where features are
insensitive to the jet definition.

\section{Setup}

As starting point for our investigation we use the results of the
differential NNLO QCD calculation presented in
Ref.~\cite{Cacciari:2015jma}, and we follow the setup chosen there. 
Specifically, we study VBF-$H$ production
\begin{equation}
pp \rightarrow H jj + X
\end{equation}
in proton-proton collisions with a center-of-mass energy of
$\sqrt{s}=13$~TeV. As parton distribution function we use the NNPDF~3.0
set at NNLO with $\alpha_S(M_Z)=0.118$
(\texttt{NNPDF30\_nnlo\_as\_0118})~\cite{Ball:2014uwa} throughout.
Bottom quarks are taken as massless and their effects are included, while
top-quark effects are excluded both as final state and in internal
lines. The CKM matrix is taken as diagonal. The Higgs boson is produced
on-shell, while for the $W$ and $Z$ boson propagators we use full
Breit-Wigner propagators. The masses, widths and the Fermi constant are set to
\begin{align}
M_W &= 80.398  \GeV \,, & \Gamma_W &= 2.141  \GeV \,, \nonumber\\
M_Z &= 91.1876 \GeV \,, & \Gamma_Z &= 2.4952 \GeV \,, \nonumber\\
M_H &= 125 \GeV \,, & G_F &= 1.16637\times10^{-5} \GeV^{-2} \,.
\label{eq:param}
\end{align}
The remaining electroweak parameters are calculated from the $W$ and $Z$
masses and $G_F$ via electroweak tree-level relations.
As scale we use
\begin{equation}
\mu^2 = \frac{M_H}2 \sqrt{\left( \frac{M_H}2 \right)^2 + p_{T,H}^2} \,.
\label{eq:scale}
\end{equation}

By default, jets are clustered from partons using the anti-$k_T$
algorithm~\cite{Cacciari:2008gp} with separation parameter $R=0.4$. 
Each event must contain at least two jets with 
\begin{align}
p_{T,j} &> 25\GeV \,.
\label{eq:cuts1}
\end{align}
To pass the VBF cuts, the two tagging jets, defined as the two jets of highest
$p_T$, must fulfill
\begin{align}
|y_j| &< 4.5 
& 
y_{j1} \times y_{j2} &< 0 
\nonumber \\
m_{j1,j2} &> 600\GeV 
& 
\Delta y_{j1,j2} &> 4.5 & 
\, ,
\label{eq:cuts2}
\end{align}
i.e.\ they must be located in opposite hemispheres, well-separated in 
rapidity, and the pair must have a large invariant mass.

To study the impact of the jet clustering, let us first consider the
effect of the algorithm. We restrict ourselves here to the family of
sequential recombination jet algorithms~\cite{Cacciari:2008gp}, with the
variants anti-$k_T$, Cambridge/Aachen (C/A), and $k_T$, which correspond
to the exponent parameter $n=-1$, $0$, and $1$, respectively.
Since the algorithm is a function of the momenta of the final-state
partons, loop effects have a strongly diminished influence as compared to 
real emission corrections.
For two-parton final states, both partons need to be identified as jets for
the event to pass the selection cuts,
Eqs.~(\ref{eq:cuts1},\ref{eq:cuts2}). Therefore, from the rapidity
separation cut it follows that the separation of the two partons is
at least as large, $R_{jj}>4.5$. This is well above any reasonable
values for the $R$ parameter of the jet algorithm, so for two-parton
kinematics the choice of $R$ has no effect. As a result, the two-loop 
contributions to the NNLO cross section do not influence jet-algorithm- 
or $R$-dependence. Three-parton final states exhibit a
dependence on $R$, which determines whether the third parton is clustered 
with either of the leading partons. However, there is as yet no dependence
on the exponent, $n$, since, with at most one parton pair available for a jet,
the $p_T$-ordering of partons has no consequence. Algorithm dependence 
only enters with four-parton events. When both additional partons are 
inside the cone of
one of the leading partons, the order of recombination can become
relevant and give different results. We note that this dependence on the
parameters $R$ and $n$ is shifted by one compared to the general case of
$X+1$~jet cross sections, where already for three-parton events both 
parameters are relevant.
This is a consequence of the rapidity-separation cut in VBF processes,
which enforces a very large $R$ separation between at least two partons.

Analogous to the study of jet shapes in NNLO DIS cross sections in 
Ref.~\cite{Kauer:1999px}, which was performed with an NLO $ep\to jj+X$ code, 
to study jet definition effects in NNLO VBF cross sections, it is 
sufficient to use a sample of VBF $H$+3~jets events at NLO
QCD~\cite{Figy:2007kv,Campanario:2013fsa}. These include the necessary 
interference of tree level and 1-loop contributions for 3-parton final states 
and the 4-parton double-real parts of the VBF-$H$ NNLO QCD
calculation. 
We generate these events with process ID 110 in
VBFNLO~\cite{Arnold:2008rz,Baglio:2014uba,VBFNLO}, using the parameter
settings given in Eqs.~(\ref{eq:param},\ref{eq:scale}) and the cuts of 
Eqs.~(\ref{eq:cuts1},\ref{eq:cuts2}).

Our starting point, the NNLO QCD calculation of
Ref.~\cite{Cacciari:2015jma}, already contains these contributions for a
specific value of $R$ and $n$, namely $R=0.4$ and $n=-1$
(anti-$k_T$). Therefore, to go to different values we can use the
$H$+3 and $H$+4~parton integrals which contribute to the calculation of the NLO
$Hjjj$ cross section to define correction terms $\Delta(R,n)$ such that
\begin{equation}
\di \sigma_{Hjj}^{\text{NNLO}}(R,n) 
  = \di \sigma_{Hjj}^{\text{NNLO}}(R{=}0.4,n{=}{-}1) 
    \underbrace{
    - \di \sigma_{H3+}^{\text{NLO}}(R{=}0.4,n{=}{-}1)
    + \di \sigma_{H3+}^{\text{NLO}}(R,n)
    }_{ = \Delta(R,n) }
\,.
\label{eq:delta}
\end{equation}
We subtract the contribution for the base values from the differential
cross section and add back the same terms for the desired new values.
The $\di \sigma_{H3+}^{\text{NLO}}$ denote the sum of 3-parton and 4-parton 
contributions which, when clustered with 3-jet cuts, would give the NLO $Hjjj$
cross section. However now, after clustering with parameters $(R,n)$, 
only 2 hard jets are required by the cut function.\footnote{Here the cut 
function is defined as the step-function, multiplying the squared matrix 
elements, which specifies whether a phase space configuration satisfies 
all cuts and belongs to a specific histogram bin.} 
The full NNLO differential 
cross section would contain additional 2-parton contributions (two-loop virtual
terms, subtraction terms for double soft or collinear 4-parton configurations,
subtraction terms for virtual soft or collinear terms, finite collinear pieces
etc.), however any 2-parton contributions are independent of $(R,n)$ and
therefore exactly cancel in the difference $\Delta(R,n)$. For 3- and 4-parton
configurations the two terms of $\Delta(R,n)$ will in general have a 
different
kinematic structure for the jets. So the VBF cuts apply individually to each of the
two terms in $\Delta(R,n)$. Similarly, when histogramming the
events, the terms might go into different bins of the distribution.

To better understand why the subtraction terms of the NLO $Hjjj$ cross 
section are
also sufficient for the calculation of $\Delta(R,n)$, let us consider the
phase space regions of non-vanishing $\Delta(R,n)$. For each 3- or 4-parton
configuration passing the VBF cuts we can identify the two leading (highest
$p_T$) hard partons, which are at the center of the two tagging jets, by running
the clustering algorithm. A non-vanishing $\Delta(R,n)$ requires at least one
hard additional parton with separation from one of the leading partons (or 
pre-cluster of partons) in the interval $(0.4,R)$. If the clusters for
$(R=0.4,n=-1)$ and $(R,n)$ differ only by sufficiently soft partons (or not at
all), both jet parameter choices will yield practically the same jets which
end up in identical histogram bins, i.e.\ one gets $\Delta(R,n)=0$. 
A well separated hard additional parton gives a finite
contribution to the integral of 3-parton configurations, and in a 4-parton
configuration the usual subtraction term for a soft or collinear fourth parton
will render the integral over this last parton finite.
 
Technically, we first generate weighted events with no restrictions on
the phase-space of the partons, both for three-parton kinematics,
comprising the Born and virtual contributions of the VBF $H$+3~jets
cross section, and for four-parton kinematics for the real-emission
part. Then for all required $(R,n)$ combinations the jet clustering
algorithm is applied to the event. If it passes the jet cuts of 
Eqs.~(\ref{eq:cuts1},\ref{eq:cuts2}), the weight of the event is booked
in the corresponding histogram bin, taking into account the minus sign
for the middle term in Eq.~\eqref{eq:delta}. To steer the Monte-Carlo
integration, we finally check if any $(R,n)$ combination has passed all
cuts. If not, the event weight is set to zero. Otherwise, in order to 
improve Monte Carlo convergence, we weight the
event with the largest differences of $m_{jj}$ and $\Delta y_{jj}$
between all considered $(R,n)$ and the reference value of
$(R{=}0.4,n{=}{-}1)$. 
For example, in a 2-jet configuration where three of the partons are
very collinear and are always clustered into the same jet, the weight
factor would become exactly zero, corresponding to the fact that in that
case also $\Delta(R,n)$ vanishes. Similarly, if one of the partons
becomes soft, its effect on $m_{jj}$ and $\Delta y_{jj}$ becomes small,
so that in the infrared limit the weight factor also approaches zero.
Note that this factor is only used to steer the integration and does
not enter the physical weights used for the histograms.

An identity similar to Eq.~\eqref{eq:delta} relates the NLO result for 
arbitrary values of $R$ with the NLO result for $R=0.4$ and the LO matrix 
elements for VBF $H$+3~jets production,
$
\di \sigma_{Hjj}^{\text{NLO}}(R) 
  = \di \sigma_{Hjj}^{\text{NLO}}(R{=}0.4) 
    - \di \sigma_{H3}^{\text{LO}}(R{=}0.4)
    + \di \sigma_{H3}^{\text{LO}}(R)
$. We have checked that the differential cross section obtained in this
way agrees with a calculation where the jet clustering radius has been
explicitly set to $R$. The level of agreement is better than one per mill.
To verify our setup, we have also cross-checked our results for the
integrated and differential LO and NLO cross sections generated by
VBFNLO with the results from Ref.~\cite{Cacciari:2015jma} and find that
they are consistent within the statistical errors from Monte-Carlo
integration.

\section{Results}

\begin{figure}[b]
\includegraphics{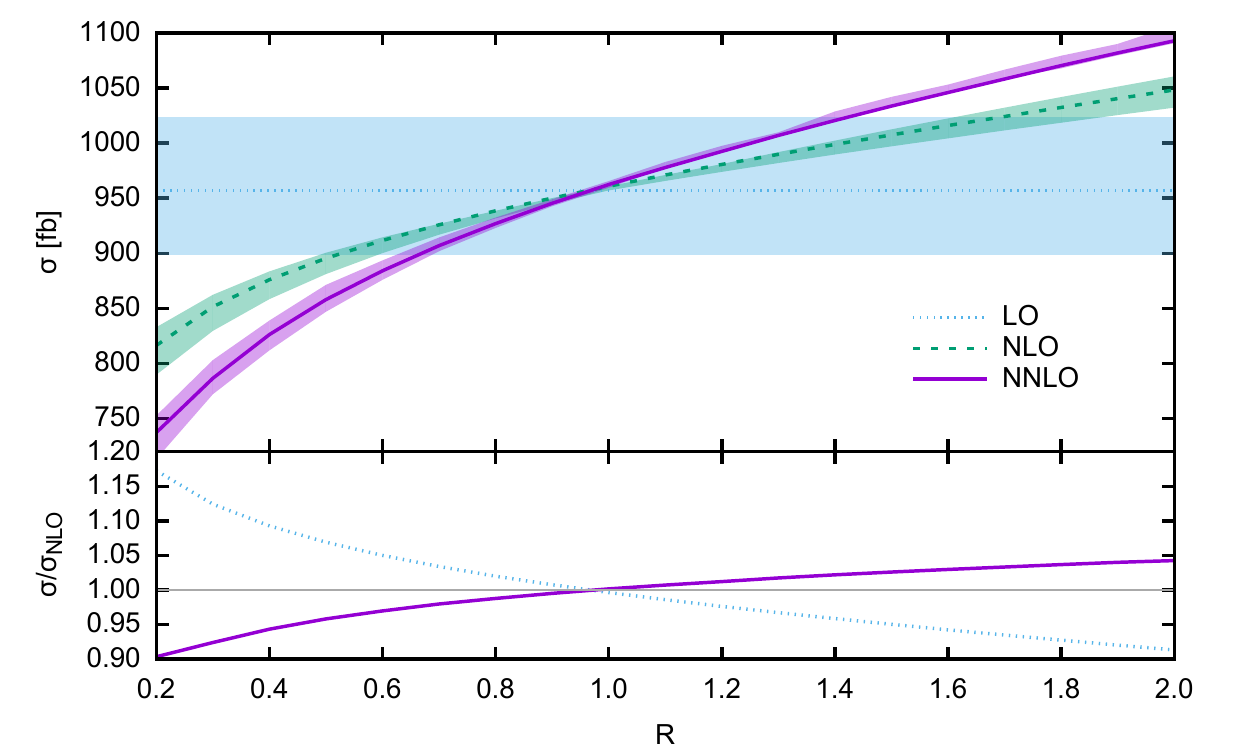}
\caption{Integrated cross section for VBF-$H$ production as function of
the jet clustering radius $R$ using the setup and cuts of
Eqs.~(\ref{eq:param},\ref{eq:scale},\ref{eq:cuts1},\ref{eq:cuts2}). 
  The respective bands arise from a scale variation by a factor $[0.5;2]$ around
the central scale $\mu$. 
The numerical value of the NNLO cross section at $R=0.4$ is taken from
Ref.~\cite{Cacciari:2015jma}.}
\label{fig:int}
\end{figure}

In Fig.~\ref{fig:int} we first show the integrated cross section as a
function of the jet clustering radius $R$. At LO, where only two partons 
are available, the cross section is
independent of the value of $R$, as discussed above. 
The
NLO cross section then exhibits a dependence on $R$, leading to an 8.5\%
reduction at the reference value $R=0.4$. Going to larger values, the
NLO cross section rises, at about $R=1.0$ it coincides with the LO
value, and surpasses it when $R$ becomes even larger. Going one order
higher in the perturbative expansion, the NNLO cross section exhibits an
even stronger $R$ dependence. 
At the reference value, $R=0.4$, 
a further 6\% reduction of the cross section compared
to NLO takes place. Again around $R=1.0$, the NNLO cross section agrees
with both the LO and NLO results. Therefore, we choose $R=1.0$ as a 
matching point in the following, when studying differential distributions. 
For all three curves, we also show the effect of varying the scale by a
factor between $0.5$ and $2$ around the central scale given in
Eq.~\eqref{eq:scale}, indicated by the underlying band. The sizable
scale dependence of the LO results is significantly reduced when going
to NLO, while going one order further to NNLO yields only a small
additional reduction. For all combinations, the scale variation bands
are not overlapping at the reference value. We also note that for NLO,
the position of the plateau coincides well with the central scale choice
when taking the matching point, $R=1.0$.

\begin{figure}
\includegraphics[width=\textwidth]{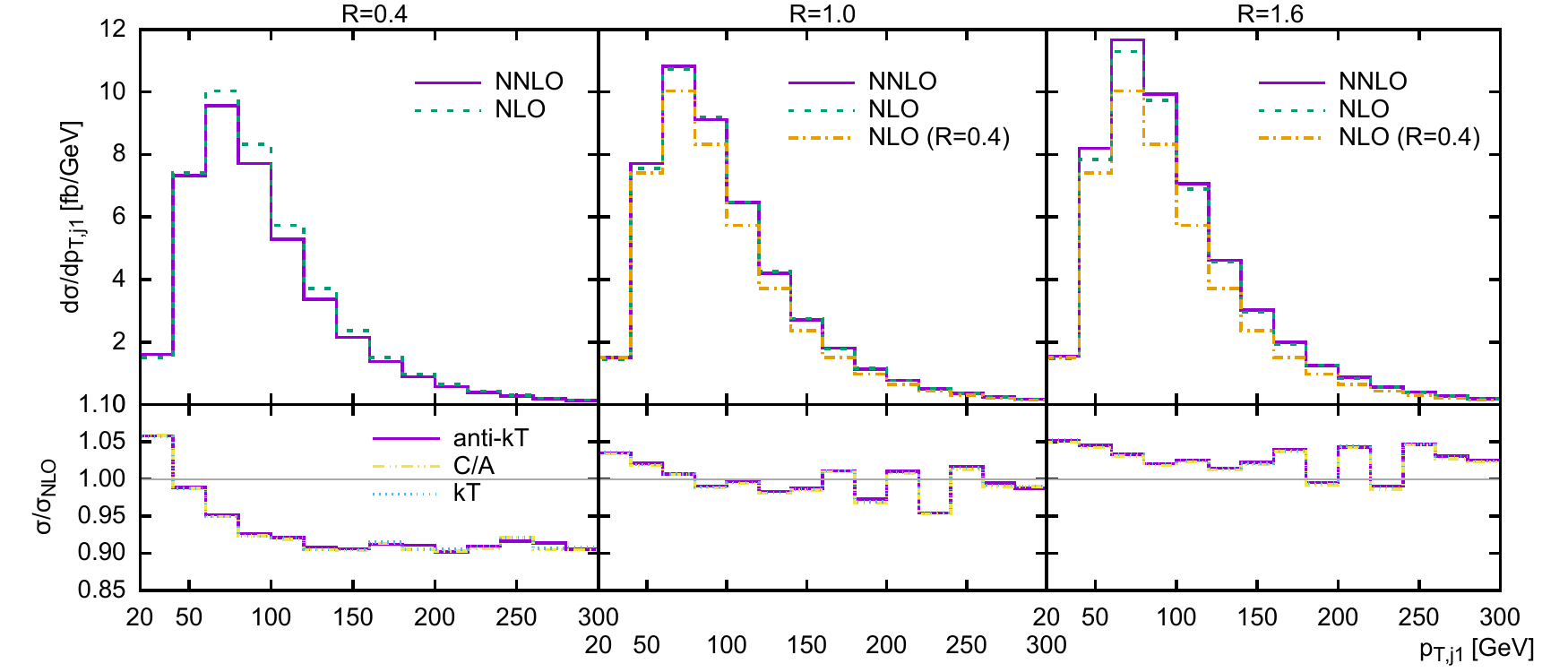}
\caption{Transverse momentum distribution of the leading jet. Results
are shown for $R$ values of $0.4$ (\textit{left}), $1.0$
(\textit{center}) and $1.6$ (\textit{right}).}
\label{fig:ptj1}
\end{figure}
Looking at distributions, we first
consider the transverse momentum of the leading jet, shown in
Fig.~\ref{fig:ptj1}. The left panel shows the results for $R=0.4$, where
the NNLO curve has been taken from Ref.~\cite{Cacciari:2015jma}. The
center panel uses the matching point with an $R$ value of $1.0$, and in
the right panel we present results with $R=1.6$ for comparison. Using
the reference value $R=0.4$, we see a reduction of the differential cross
section over most of the shown range, which is approximately a constant
factor for transverse momenta larger than 100~GeV. For smaller
transverse momenta, the corrections decrease and in the first bin the
additional NNLO effects give a positive contribution to the cross
section. Employing our matching value, $R=1.0$, we see that for
jet transverse momenta above 80~GeV, the NLO and NNLO results
roughly agree within the statistical errors. For comparison, we have
again plotted the $R{=}0.4$ NLO curve in the upper panel. It shows 
that both the NLO and NNLO curves have shifted upwards,
consistent with the behavior seen for the integrated cross section in
Fig.~\ref{fig:int}. Only in the first two bins there is still a
significant positive correction, but with a smaller relative size than
for $R=0.4$. In the right panel with $R=1.6$, we finally see that the
NNLO contributions are always positive. In the lower
ratio panels, we additionally show the effect from using the
Cambridge/Aachen and $k_T$ clustering algorithm instead of our default
choice of anti-$k_T$. The effects due to different cluster algorithms 
are tiny. We have checked that this also holds for all other
distributions. Therefore, we will show further distributions only for the 
anti-$k_T$ algorithm.

\begin{figure}
\includegraphics[width=\textwidth]{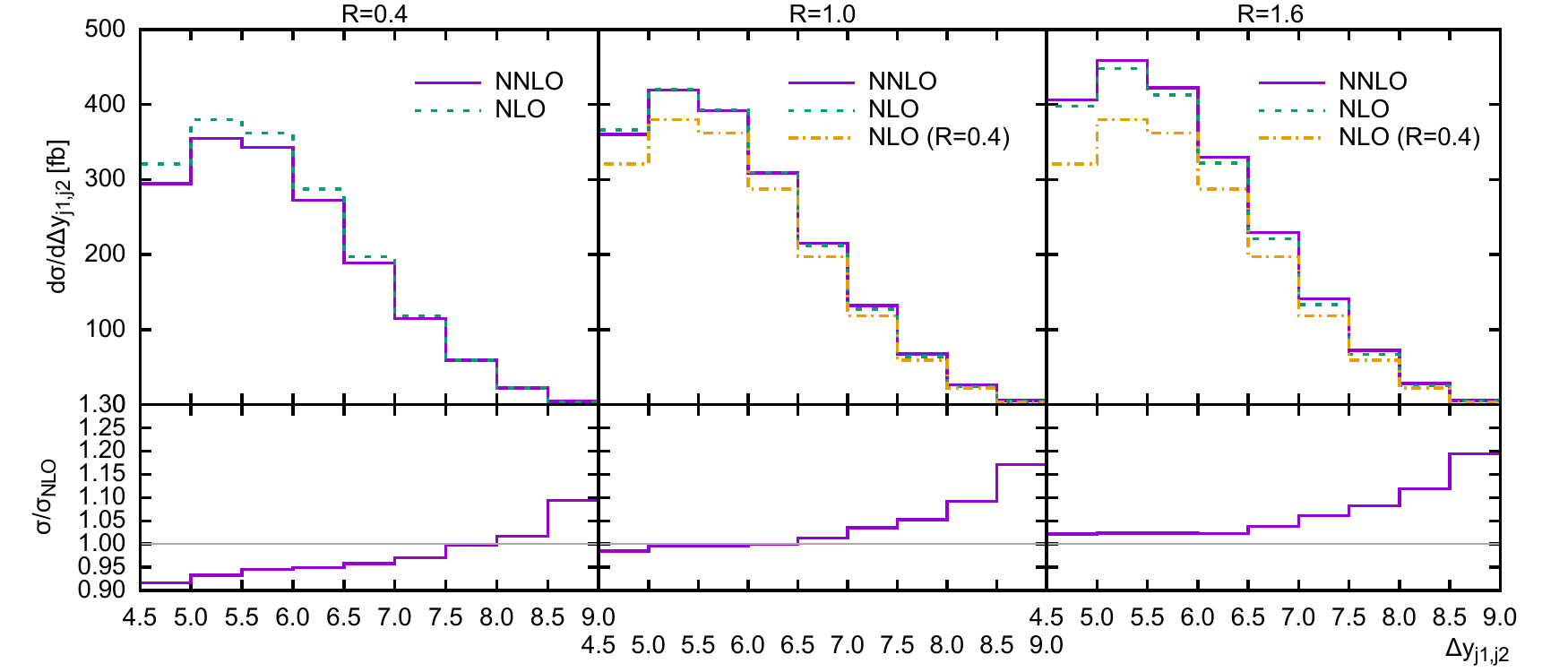}
\caption{Rapidity difference of the two tagging jets. Results
are shown for $R$ values of $0.4$ (\textit{left}), $1.0$
(\textit{center}) and $1.6$ (\textit{right}).}
\label{fig:dyjj}
\end{figure}
In Fig.~\ref{fig:dyjj}, we show the distribution of the rapidity
difference of the two tagging jets, again for the three jet clustering
radii of $R=0.4$, $1.0$ and $1.6$. For the reference value, the ratio
of NNLO over NLO cross sections is below one for values close to the
lower cut at $\Delta y_{j1,j2}=4.5$. They then become gradually larger
until a large positive correction is reached at the upper limit of $9$,
which is given by the requirement that the absolute value of the jet
rapidity is below $4.5$. Moving to the larger $R=1.0$ clustering, the
additional NNLO corrections become small up to a rapidity difference of
about 7. Only for larger values do we see significant remaining
contributions, which are positive in size. Thereby, the phase-space
region of large rapidity difference and small transverse momenta of the
leading jet is connected. 
As the cut on the invariant mass of the two tagging jets requires the
jets to be very energetic, a small transverse momentum causes a large
rapidity of each jet, and thus a large rapidity difference.
The distribution for $R=1.6$ exhibits a behavior similar to the one for
$R=1.0$, namely a constant correction factor for lower rapidity
differences with a rise towards larger ones. Due to the larger jet
radius, the correction factor is larger than one throughout.

\begin{figure}
\mbox{}
\hfill
\includegraphics[width=0.42857\textwidth]{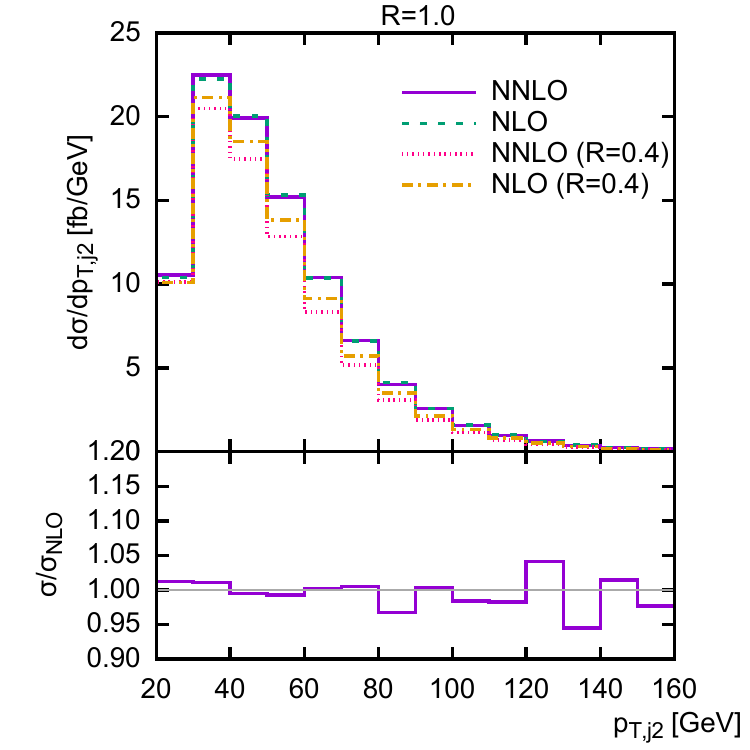}
\hfill
\includegraphics[width=0.42857\textwidth]{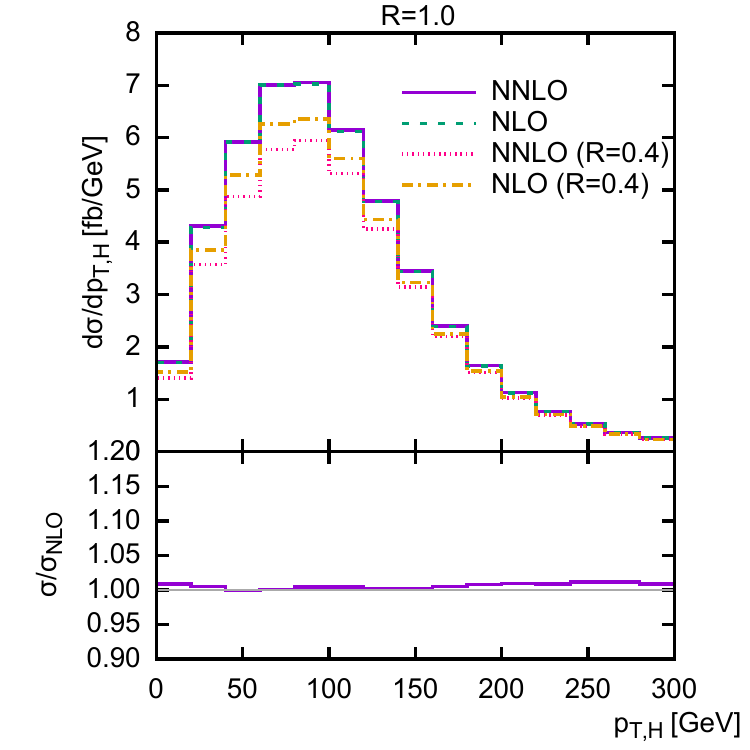}
\hfill
\mbox{}
\caption{Transverse momentum distributions of the second jet (\textit{left})
and of the Higgs boson (\textit{right}).}
\label{fig:ptj2ptH}
\end{figure}
In Fig.~\ref{fig:ptj2ptH} we finally show two more distributions, namely
the transverse momentum of the second jet on the left and the
transverse momentum of the Higgs boson on the right, showing only
results for the matching value of $R=1.0$. In both cases, the larger jet
clustering radius reduces the size of the NNLO corrections
to very small values over the whole range of the distribution. 
In contrast, at $R=0.4$ the ratio of NNLO over NLO results for these 
two distributions ranges between 0.9 and 1.0~\cite{Cacciari:2015jma} 
and thus shows sizable NNLO corrections. Obviously, these large corrections 
are due to the wider energy flow inside the tagging jets, which is captured 
to a larger degree by the $R=1.0$ jets than by the narrow $R=0.4$ jets used 
in many LHC analyses.

\section{Discussion and Conclusions}
The NNLO cross section for VBF-$Hjj$ production shows a remarkably strong
dependence on the jet definition, in particular on the jet radius in the
rapidity-azimuthal angle-plane. Starting with the results of the NNLO 
calculation for a given jet algorithm, we use the NLO QCD calculation 
of VBF $H$+3~jets production to calculate the NNLO cross section at different
values of both the jet clustering radius $R$ and the momentum exponent, 
corresponding to the choices of anti-$k_T$, Cambridge/Aachen and $k_T$. 

We find that a large jet clustering radius of $R=1.0$ leads to small NNLO
corrections to the fiducial VBF cross section (within the VBF cuts of
Eqs.(\ref{eq:cuts2})). This also holds for most differential
distributions. Relevant NNLO 
corrections remain in a phase-space region where the transverse momenta
of both jets are small, and therefore the rapidity separation between
them becomes large. Effects from choosing different jet clustering
exponents are tiny.

The strong $R$-dependence of differential distributions and fiducial cross
sections can be explained by the wider energy flow within quark jets at NNLO
QCD, which was first discussed quantitatively for DIS jets in
$ep$-scattering~\cite{Kauer:1999px}: a larger jet radius 
captures  a larger fraction of the energy of the original scattered quark and
thus leads to an increase in average jet energy. Tagging jets defined with
larger $R$ thus have larger dijet invariant mass and they more easily 
pass the $m_{jj}>600$~GeV cut of the VBF selection. The resulting increase in
fiducial cross section, exhibited in Fig.~\ref{fig:int}, is substantial,
reaching 17 percent when comparing $R=0.4$ and $R=1.0$, with a 6 percent shift
due to NNLO effects alone, a surprisingly large contribution when compared to
the considerably smaller NNLO corrections to the inclusive VBF $Hjj$ cross
section. These sizable corrections 
become intuitively understandable, however, when remembering that any
$R$-dependence is determined at one order lower in perturbation theory than
the cross section itself, i.e.\ in the NNLO cross section, $R$-dependence is
modeled with NLO accuracy only. 

The observed large corrections to jet shape and $R$-dependence, in going from
NLO to NNLO modeling, are not captured by a scale variation of NLO cross
sections. They should be treated as an additional uncertainty in any NLO cross
section calculation for processes which exhibit jets at LO already. From our 
example of VBF Higgs production, one should assign an additional, order 10\% 
uncertainty to NLO cross sections with quark jets in the final state, which 
becomes especially relevant when the scale variation of the NLO cross section 
is exceptionally small, like in VBF. For processes with gluon jets at LO,
these corrections are expected to be larger, because of the enhanced radiation 
from gluons due to their larger color charge. Also, cross sections with steeper 
jet transverse momentum dependence at LO should be affected more by energy
flow corrections in NNLO calculations than the VBF distributions, which have
relatively mild $p_T$-dependence of the tagging jets.

We found very small NNLO corrections for fiducial VBF cross sections and
distributions for tagging jets defined with a radius $R=1.0$. This does not
mean, of course, that analyses at the LHC for VBF processes should be
performed with such fat jets, since contributions from underlying event or
pile-up would lead to additional large corrections, which may well induce
higher cross section uncertainties than N3LO effects, which are still missing
in the discussion above. 
Such investigations go beyond the aim and scope of the present
paper, however.

\begin{acknowledgments}
We would like to thank the authors of Ref.~\cite{Cacciari:2015jma} for
providing the raw histogram data of their results. We gratefully acknowledge 
support from ``BMBF Verbundforschung Teilchenphysik`` under grant
number 05H15VKCCA.
\end{acknowledgments}

\bibliographystyle{apsrev4-1}
\bibliography{papers}

\begin{thebibliography}{40}%
\makeatletter
\providecommand \@ifxundefined [1]{%
 \@ifx{#1\undefined}
}%
\providecommand \@ifnum [1]{%
 \ifnum #1\expandafter \@firstoftwo
 \else \expandafter \@secondoftwo
 \fi
}%
\providecommand \@ifx [1]{%
 \ifx #1\expandafter \@firstoftwo
 \else \expandafter \@secondoftwo
 \fi
}%
\providecommand \natexlab [1]{#1}%
\providecommand \enquote  [1]{``#1''}%
\providecommand \bibnamefont  [1]{#1}%
\providecommand \bibfnamefont [1]{#1}%
\providecommand \citenamefont [1]{#1}%
\providecommand \href@noop [0]{\@secondoftwo}%
\providecommand \href [0]{\begingroup \@sanitize@url \@href}%
\providecommand \@href[1]{\@@startlink{#1}\@@href}%
\providecommand \@@href[1]{\endgroup#1\@@endlink}%
\providecommand \@sanitize@url [0]{\catcode `\\12\catcode `\$12\catcode
  `\&12\catcode `\#12\catcode `\^12\catcode `\_12\catcode `\%12\relax}%
\providecommand \@@startlink[1]{}%
\providecommand \@@endlink[0]{}%
\providecommand \url  [0]{\begingroup\@sanitize@url \@url }%
\providecommand \@url [1]{\endgroup\@href {#1}{\urlprefix }}%
\providecommand \urlprefix  [0]{URL }%
\providecommand \Eprint [0]{\href }%
\providecommand \doibase [0]{http://dx.doi.org/}%
\providecommand \selectlanguage [0]{\@gobble}%
\providecommand \bibinfo  [0]{\@secondoftwo}%
\providecommand \bibfield  [0]{\@secondoftwo}%
\providecommand \translation [1]{[#1]}%
\providecommand \BibitemOpen [0]{}%
\providecommand \bibitemStop [0]{}%
\providecommand \bibitemNoStop [0]{.\EOS\space}%
\providecommand \EOS [0]{\spacefactor3000\relax}%
\providecommand \BibitemShut  [1]{\csname bibitem#1\endcsname}%
\let\auto@bib@innerbib\@empty
\bibitem [{\citenamefont {Cacciari}\ \emph {et~al.}(2015)\citenamefont
  {Cacciari}, \citenamefont {Dreyer}, \citenamefont {Karlberg}, \citenamefont
  {Salam},\ and\ \citenamefont {Zanderighi}}]{Cacciari:2015jma}%
  \BibitemOpen
  \bibfield  {author} {\bibinfo {author} {\bibfnamefont {M.}~\bibnamefont
  {Cacciari}}, \bibinfo {author} {\bibfnamefont {F.~A.}\ \bibnamefont
  {Dreyer}}, \bibinfo {author} {\bibfnamefont {A.}~\bibnamefont {Karlberg}},
  \bibinfo {author} {\bibfnamefont {G.~P.}\ \bibnamefont {Salam}}, \ and\
  \bibinfo {author} {\bibfnamefont {G.}~\bibnamefont {Zanderighi}},\ }\href
  {\doibase 10.1103/PhysRevLett.115.082002} {\bibfield  {journal} {\bibinfo
  {journal} {Phys. Rev. Lett.}\ }\textbf {\bibinfo {volume} {115}},\ \bibinfo
  {pages} {082002} (\bibinfo {year} {2015})},\ \Eprint
  {http://arxiv.org/abs/1506.02660} {arXiv:1506.02660 [hep-ph]} \BibitemShut
  {NoStop}%
\bibitem [{\citenamefont {Aad}\ \emph {et~al.}(2012)\citenamefont {Aad} \emph
  {et~al.}}]{Aad:2012tfa}%
  \BibitemOpen
  \bibfield  {author} {\bibinfo {author} {\bibfnamefont {G.}~\bibnamefont
  {Aad}} \emph {et~al.} (\bibinfo {collaboration} {ATLAS}),\ }\href {\doibase
  10.1016/j.physletb.2012.08.020} {\bibfield  {journal} {\bibinfo  {journal}
  {Phys. Lett.}\ }\textbf {\bibinfo {volume} {B716}},\ \bibinfo {pages} {1}
  (\bibinfo {year} {2012})},\ \Eprint {http://arxiv.org/abs/1207.7214}
  {arXiv:1207.7214 [hep-ex]} \BibitemShut {NoStop}%
\bibitem [{\citenamefont {Chatrchyan}\ \emph {et~al.}(2012)\citenamefont
  {Chatrchyan} \emph {et~al.}}]{Chatrchyan:2012xdj}%
  \BibitemOpen
  \bibfield  {author} {\bibinfo {author} {\bibfnamefont {S.}~\bibnamefont
  {Chatrchyan}} \emph {et~al.} (\bibinfo {collaboration} {CMS}),\ }\href
  {\doibase 10.1016/j.physletb.2012.08.021} {\bibfield  {journal} {\bibinfo
  {journal} {Phys. Lett.}\ }\textbf {\bibinfo {volume} {B716}},\ \bibinfo
  {pages} {30} (\bibinfo {year} {2012})},\ \Eprint
  {http://arxiv.org/abs/1207.7235} {arXiv:1207.7235 [hep-ex]} \BibitemShut
  {NoStop}%
\bibitem [{\citenamefont {Cahn}\ and\ \citenamefont
  {Dawson}(1984)}]{Cahn:1983ip}%
  \BibitemOpen
  \bibfield  {author} {\bibinfo {author} {\bibfnamefont {R.~N.}\ \bibnamefont
  {Cahn}}\ and\ \bibinfo {author} {\bibfnamefont {S.}~\bibnamefont {Dawson}},\
  }\href {\doibase 10.1016/0370-2693(84)91180-8} {\bibfield  {journal}
  {\bibinfo  {journal} {Phys. Lett.}\ }\textbf {\bibinfo {volume} {B136}},\
  \bibinfo {pages} {196} (\bibinfo {year} {1984})},\ \bibinfo {note} {[Erratum:
  Phys. Lett.B138,464(1984)]}\BibitemShut {NoStop}%
\bibitem [{\citenamefont {Dawson}(1985)}]{Dawson:1984gx}%
  \BibitemOpen
  \bibfield  {author} {\bibinfo {author} {\bibfnamefont {S.}~\bibnamefont
  {Dawson}},\ }\href {\doibase 10.1016/0550-3213(85)90038-0} {\bibfield
  {journal} {\bibinfo  {journal} {Nucl. Phys.}\ }\textbf {\bibinfo {volume}
  {B249}},\ \bibinfo {pages} {42} (\bibinfo {year} {1985})}\BibitemShut
  {NoStop}%
\bibitem [{\citenamefont {Duncan}\ \emph {et~al.}(1986)\citenamefont {Duncan},
  \citenamefont {Kane},\ and\ \citenamefont {Repko}}]{Duncan:1985vj}%
  \BibitemOpen
  \bibfield  {author} {\bibinfo {author} {\bibfnamefont {M.~J.}\ \bibnamefont
  {Duncan}}, \bibinfo {author} {\bibfnamefont {G.~L.}\ \bibnamefont {Kane}}, \
  and\ \bibinfo {author} {\bibfnamefont {W.~W.}\ \bibnamefont {Repko}},\ }\href
  {\doibase 10.1016/0550-3213(86)90234-8} {\bibfield  {journal} {\bibinfo
  {journal} {Nucl. Phys.}\ }\textbf {\bibinfo {volume} {B272}},\ \bibinfo
  {pages} {517} (\bibinfo {year} {1986})}\BibitemShut {NoStop}%
\bibitem [{\citenamefont {Cahn}\ \emph {et~al.}(1987)\citenamefont {Cahn},
  \citenamefont {Ellis}, \citenamefont {Kleiss},\ and\ \citenamefont
  {Stirling}}]{Cahn:1986zv}%
  \BibitemOpen
  \bibfield  {author} {\bibinfo {author} {\bibfnamefont {R.~N.}\ \bibnamefont
  {Cahn}}, \bibinfo {author} {\bibfnamefont {S.~D.}\ \bibnamefont {Ellis}},
  \bibinfo {author} {\bibfnamefont {R.}~\bibnamefont {Kleiss}}, \ and\ \bibinfo
  {author} {\bibfnamefont {W.~J.}\ \bibnamefont {Stirling}},\ }\href {\doibase
  10.1103/PhysRevD.35.1626} {\bibfield  {journal} {\bibinfo  {journal} {Phys.
  Rev.}\ }\textbf {\bibinfo {volume} {D35}},\ \bibinfo {pages} {1626} (\bibinfo
  {year} {1987})}\BibitemShut {NoStop}%
\bibitem [{\citenamefont {Kleiss}\ and\ \citenamefont
  {Stirling}(1988)}]{Kleiss:1987cj}%
  \BibitemOpen
  \bibfield  {author} {\bibinfo {author} {\bibfnamefont {R.}~\bibnamefont
  {Kleiss}}\ and\ \bibinfo {author} {\bibfnamefont {W.~J.}\ \bibnamefont
  {Stirling}},\ }\href {\doibase 10.1016/0370-2693(88)91135-5} {\bibfield
  {journal} {\bibinfo  {journal} {Phys. Lett.}\ }\textbf {\bibinfo {volume}
  {B200}},\ \bibinfo {pages} {193} (\bibinfo {year} {1988})}\BibitemShut
  {NoStop}%
\bibitem [{\citenamefont {Barger}\ \emph {et~al.}(1988)\citenamefont {Barger},
  \citenamefont {Han},\ and\ \citenamefont {Phillips}}]{Barger:1988mr}%
  \BibitemOpen
  \bibfield  {author} {\bibinfo {author} {\bibfnamefont {V.~D.}\ \bibnamefont
  {Barger}}, \bibinfo {author} {\bibfnamefont {T.}~\bibnamefont {Han}}, \ and\
  \bibinfo {author} {\bibfnamefont {R.~J.~N.}\ \bibnamefont {Phillips}},\
  }\href {\doibase 10.1103/PhysRevD.37.2005} {\bibfield  {journal} {\bibinfo
  {journal} {Phys. Rev.}\ }\textbf {\bibinfo {volume} {D37}},\ \bibinfo {pages}
  {2005} (\bibinfo {year} {1988})}\BibitemShut {NoStop}%
\bibitem [{\citenamefont {Butterworth}\ \emph {et~al.}(2002)\citenamefont
  {Butterworth}, \citenamefont {Cox},\ and\ \citenamefont
  {Forshaw}}]{Butterworth:2002tt}%
  \BibitemOpen
  \bibfield  {author} {\bibinfo {author} {\bibfnamefont {J.~M.}\ \bibnamefont
  {Butterworth}}, \bibinfo {author} {\bibfnamefont {B.~E.}\ \bibnamefont
  {Cox}}, \ and\ \bibinfo {author} {\bibfnamefont {J.~R.}\ \bibnamefont
  {Forshaw}},\ }\href {\doibase 10.1103/PhysRevD.65.096014} {\bibfield
  {journal} {\bibinfo  {journal} {Phys. Rev.}\ }\textbf {\bibinfo {volume}
  {D65}},\ \bibinfo {pages} {096014} (\bibinfo {year} {2002})},\ \Eprint
  {http://arxiv.org/abs/hep-ph/0201098} {arXiv:hep-ph/0201098 [hep-ph]}
  \BibitemShut {NoStop}%
\bibitem [{\citenamefont {Dittmaier}\ \emph {et~al.}(2011)\citenamefont
  {Dittmaier} \emph {et~al.}}]{Dittmaier:2011ti}%
  \BibitemOpen
  \bibfield  {author} {\bibinfo {author} {\bibfnamefont {S.}~\bibnamefont
  {Dittmaier}} \emph {et~al.} (\bibinfo {collaboration} {LHC Higgs Cross
  Section Working Group}),\ }\href {\doibase 10.5170/CERN-2011-002} {\
  (\bibinfo {year} {2011}),\ 10.5170/CERN-2011-002},\ \Eprint
  {http://arxiv.org/abs/1101.0593} {arXiv:1101.0593 [hep-ph]} \BibitemShut
  {NoStop}%
\bibitem [{\citenamefont {Dittmaier}\ \emph {et~al.}(2012)\citenamefont
  {Dittmaier} \emph {et~al.}}]{Dittmaier:2012vm}%
  \BibitemOpen
  \bibfield  {author} {\bibinfo {author} {\bibfnamefont {S.}~\bibnamefont
  {Dittmaier}} \emph {et~al.},\ }\href {\doibase 10.5170/CERN-2012-002} {\
  (\bibinfo {year} {2012}),\ 10.5170/CERN-2012-002},\ \Eprint
  {http://arxiv.org/abs/1201.3084} {arXiv:1201.3084 [hep-ph]} \BibitemShut
  {NoStop}%
\bibitem [{\citenamefont {Andersen}\ \emph {et~al.}(2013)\citenamefont
  {Andersen} \emph {et~al.}}]{Heinemeyer:2013tqa}%
  \BibitemOpen
  \bibfield  {author} {\bibinfo {author} {\bibfnamefont {J.~R.}\ \bibnamefont
  {Andersen}} \emph {et~al.} (\bibinfo {collaboration} {LHC Higgs Cross Section
  Working Group}),\ }\href {\doibase 10.5170/CERN-2013-004} {\  (\bibinfo
  {year} {2013}),\ 10.5170/CERN-2013-004},\ \Eprint
  {http://arxiv.org/abs/1307.1347} {arXiv:1307.1347 [hep-ph]} \BibitemShut
  {NoStop}%
\bibitem [{\citenamefont {Rauch}(2016)}]{Rauch:2016pai}%
  \BibitemOpen
  \bibfield  {author} {\bibinfo {author} {\bibfnamefont {M.}~\bibnamefont
  {Rauch}},\ }\href@noop {} {\  (\bibinfo {year} {2016})},\ \Eprint
  {http://arxiv.org/abs/1610.08420} {arXiv:1610.08420 [hep-ph]} \BibitemShut
  {NoStop}%
\bibitem [{\citenamefont {Zeppenfeld}(1999)}]{Zeppenfeld:1999yd}%
  \BibitemOpen
  \bibfield  {author} {\bibinfo {author} {\bibfnamefont {D.}~\bibnamefont
  {Zeppenfeld}},\ }in\ \href
  {http://alice.cern.ch/format/showfull?sysnb=0304571} {\emph {\bibinfo
  {booktitle} {{Neutrinos in physics and astrophysics from 10**(-33) to 10**28
  CM. Proceedings, Conference, TASI'98, Boulder, USA, June 1-26, 1998}}}}\
  (\bibinfo {year} {1999})\ pp.\ \bibinfo {pages} {303--350},\ \Eprint
  {http://arxiv.org/abs/hep-ph/9902307} {arXiv:hep-ph/9902307 [hep-ph]}
  \BibitemShut {NoStop}%
\bibitem [{\citenamefont {Plehn}\ \emph {et~al.}(2000)\citenamefont {Plehn},
  \citenamefont {Rainwater},\ and\ \citenamefont {Zeppenfeld}}]{Plehn:1999xi}%
  \BibitemOpen
  \bibfield  {author} {\bibinfo {author} {\bibfnamefont {T.}~\bibnamefont
  {Plehn}}, \bibinfo {author} {\bibfnamefont {D.~L.}\ \bibnamefont
  {Rainwater}}, \ and\ \bibinfo {author} {\bibfnamefont {D.}~\bibnamefont
  {Zeppenfeld}},\ }\href {\doibase 10.1103/PhysRevD.61.093005} {\bibfield
  {journal} {\bibinfo  {journal} {Phys. Rev.}\ }\textbf {\bibinfo {volume}
  {D61}},\ \bibinfo {pages} {093005} (\bibinfo {year} {2000})},\ \Eprint
  {http://arxiv.org/abs/hep-ph/9911385} {arXiv:hep-ph/9911385 [hep-ph]}
  \BibitemShut {NoStop}%
\bibitem [{\citenamefont {Aad}\ \emph {et~al.}(2015)\citenamefont {Aad} \emph
  {et~al.}}]{Aad:2015vsa}%
  \BibitemOpen
  \bibfield  {author} {\bibinfo {author} {\bibfnamefont {G.}~\bibnamefont
  {Aad}} \emph {et~al.} (\bibinfo {collaboration} {ATLAS}),\ }\href {\doibase
  10.1007/JHEP04(2015)117} {\bibfield  {journal} {\bibinfo  {journal} {JHEP}\
  }\textbf {\bibinfo {volume} {04}},\ \bibinfo {pages} {117} (\bibinfo {year}
  {2015})},\ \Eprint {http://arxiv.org/abs/1501.04943} {arXiv:1501.04943
  [hep-ex]} \BibitemShut {NoStop}%
\bibitem [{\citenamefont {Chatrchyan}\ \emph {et~al.}(2014)\citenamefont
  {Chatrchyan} \emph {et~al.}}]{Chatrchyan:2014nva}%
  \BibitemOpen
  \bibfield  {author} {\bibinfo {author} {\bibfnamefont {S.}~\bibnamefont
  {Chatrchyan}} \emph {et~al.} (\bibinfo {collaboration} {CMS}),\ }\href
  {\doibase 10.1007/JHEP05(2014)104} {\bibfield  {journal} {\bibinfo  {journal}
  {JHEP}\ }\textbf {\bibinfo {volume} {05}},\ \bibinfo {pages} {104} (\bibinfo
  {year} {2014})},\ \Eprint {http://arxiv.org/abs/1401.5041} {arXiv:1401.5041
  [hep-ex]} \BibitemShut {NoStop}%
\bibitem [{\citenamefont {Eboli}\ and\ \citenamefont
  {Zeppenfeld}(2000)}]{Eboli:2000ze}%
  \BibitemOpen
  \bibfield  {author} {\bibinfo {author} {\bibfnamefont {O.~J.~P.}\
  \bibnamefont {Eboli}}\ and\ \bibinfo {author} {\bibfnamefont
  {D.}~\bibnamefont {Zeppenfeld}},\ }\href {\doibase
  10.1016/S0370-2693(00)01213-2} {\bibfield  {journal} {\bibinfo  {journal}
  {Phys. Lett.}\ }\textbf {\bibinfo {volume} {B495}},\ \bibinfo {pages} {147}
  (\bibinfo {year} {2000})},\ \Eprint {http://arxiv.org/abs/hep-ph/0009158}
  {arXiv:hep-ph/0009158 [hep-ph]} \BibitemShut {NoStop}%
\bibitem [{\citenamefont {Aad}\ \emph {et~al.}(2016)\citenamefont {Aad} \emph
  {et~al.}}]{Aad:2015txa}%
  \BibitemOpen
  \bibfield  {author} {\bibinfo {author} {\bibfnamefont {G.}~\bibnamefont
  {Aad}} \emph {et~al.} (\bibinfo {collaboration} {ATLAS}),\ }\href {\doibase
  10.1007/JHEP01(2016)172} {\bibfield  {journal} {\bibinfo  {journal} {JHEP}\
  }\textbf {\bibinfo {volume} {01}},\ \bibinfo {pages} {172} (\bibinfo {year}
  {2016})},\ \Eprint {http://arxiv.org/abs/1508.07869} {arXiv:1508.07869
  [hep-ex]} \BibitemShut {NoStop}%
\bibitem [{\citenamefont {Khachatryan}\ \emph {et~al.}(2017)\citenamefont
  {Khachatryan} \emph {et~al.}}]{Khachatryan:2016whc}%
  \BibitemOpen
  \bibfield  {author} {\bibinfo {author} {\bibfnamefont {V.}~\bibnamefont
  {Khachatryan}} \emph {et~al.} (\bibinfo {collaboration} {CMS}),\ }\href
  {\doibase 10.1007/JHEP02(2017)135} {\bibfield  {journal} {\bibinfo  {journal}
  {JHEP}\ }\textbf {\bibinfo {volume} {02}},\ \bibinfo {pages} {135} (\bibinfo
  {year} {2017})},\ \Eprint {http://arxiv.org/abs/1610.09218} {arXiv:1610.09218
  [hep-ex]} \BibitemShut {NoStop}%
\bibitem [{\citenamefont {Plehn}\ \emph {et~al.}(2002)\citenamefont {Plehn},
  \citenamefont {Rainwater},\ and\ \citenamefont {Zeppenfeld}}]{Plehn:2001nj}%
  \BibitemOpen
  \bibfield  {author} {\bibinfo {author} {\bibfnamefont {T.}~\bibnamefont
  {Plehn}}, \bibinfo {author} {\bibfnamefont {D.~L.}\ \bibnamefont
  {Rainwater}}, \ and\ \bibinfo {author} {\bibfnamefont {D.}~\bibnamefont
  {Zeppenfeld}},\ }\href {\doibase 10.1103/PhysRevLett.88.051801} {\bibfield
  {journal} {\bibinfo  {journal} {Phys. Rev. Lett.}\ }\textbf {\bibinfo
  {volume} {88}},\ \bibinfo {pages} {051801} (\bibinfo {year} {2002})},\
  \Eprint {http://arxiv.org/abs/hep-ph/0105325} {arXiv:hep-ph/0105325 [hep-ph]}
  \BibitemShut {NoStop}%
\bibitem [{\citenamefont {Andersen}\ \emph {et~al.}(2010)\citenamefont
  {Andersen}, \citenamefont {Arnold},\ and\ \citenamefont
  {Zeppenfeld}}]{Andersen:2010zx}%
  \BibitemOpen
  \bibfield  {author} {\bibinfo {author} {\bibfnamefont {J.~R.}\ \bibnamefont
  {Andersen}}, \bibinfo {author} {\bibfnamefont {K.}~\bibnamefont {Arnold}}, \
  and\ \bibinfo {author} {\bibfnamefont {D.}~\bibnamefont {Zeppenfeld}},\
  }\href {\doibase 10.1007/JHEP06(2010)091} {\bibfield  {journal} {\bibinfo
  {journal} {JHEP}\ }\textbf {\bibinfo {volume} {06}},\ \bibinfo {pages} {091}
  (\bibinfo {year} {2010})},\ \Eprint {http://arxiv.org/abs/1001.3822}
  {arXiv:1001.3822 [hep-ph]} \BibitemShut {NoStop}%
\bibitem [{\citenamefont {Han}\ \emph {et~al.}(1992)\citenamefont {Han},
  \citenamefont {Valencia},\ and\ \citenamefont {Willenbrock}}]{Han:1992hr}%
  \BibitemOpen
  \bibfield  {author} {\bibinfo {author} {\bibfnamefont {T.}~\bibnamefont
  {Han}}, \bibinfo {author} {\bibfnamefont {G.}~\bibnamefont {Valencia}}, \
  and\ \bibinfo {author} {\bibfnamefont {S.}~\bibnamefont {Willenbrock}},\
  }\href {\doibase 10.1103/PhysRevLett.69.3274} {\bibfield  {journal} {\bibinfo
   {journal} {Phys. Rev. Lett.}\ }\textbf {\bibinfo {volume} {69}},\ \bibinfo
  {pages} {3274} (\bibinfo {year} {1992})},\ \Eprint
  {http://arxiv.org/abs/hep-ph/9206246} {arXiv:hep-ph/9206246 [hep-ph]}
  \BibitemShut {NoStop}%
\bibitem [{\citenamefont {Figy}\ \emph {et~al.}(2003)\citenamefont {Figy},
  \citenamefont {Oleari},\ and\ \citenamefont {Zeppenfeld}}]{Figy:2003nv}%
  \BibitemOpen
  \bibfield  {author} {\bibinfo {author} {\bibfnamefont {T.}~\bibnamefont
  {Figy}}, \bibinfo {author} {\bibfnamefont {C.}~\bibnamefont {Oleari}}, \ and\
  \bibinfo {author} {\bibfnamefont {D.}~\bibnamefont {Zeppenfeld}},\ }\href
  {\doibase 10.1103/PhysRevD.68.073005} {\bibfield  {journal} {\bibinfo
  {journal} {Phys. Rev.}\ }\textbf {\bibinfo {volume} {D68}},\ \bibinfo {pages}
  {073005} (\bibinfo {year} {2003})},\ \Eprint
  {http://arxiv.org/abs/hep-ph/0306109} {arXiv:hep-ph/0306109 [hep-ph]}
  \BibitemShut {NoStop}%
\bibitem [{\citenamefont {Berger}\ and\ \citenamefont
  {Campbell}(2004)}]{Berger:2004pca}%
  \BibitemOpen
  \bibfield  {author} {\bibinfo {author} {\bibfnamefont {E.~L.}\ \bibnamefont
  {Berger}}\ and\ \bibinfo {author} {\bibfnamefont {J.~M.}\ \bibnamefont
  {Campbell}},\ }\href {\doibase 10.1103/PhysRevD.70.073011} {\bibfield
  {journal} {\bibinfo  {journal} {Phys. Rev.}\ }\textbf {\bibinfo {volume}
  {D70}},\ \bibinfo {pages} {073011} (\bibinfo {year} {2004})},\ \Eprint
  {http://arxiv.org/abs/hep-ph/0403194} {arXiv:hep-ph/0403194 [hep-ph]}
  \BibitemShut {NoStop}%
\bibitem [{\citenamefont {Ciccolini}\ \emph {et~al.}(2007)\citenamefont
  {Ciccolini}, \citenamefont {Denner},\ and\ \citenamefont
  {Dittmaier}}]{Ciccolini:2007jr}%
  \BibitemOpen
  \bibfield  {author} {\bibinfo {author} {\bibfnamefont {M.}~\bibnamefont
  {Ciccolini}}, \bibinfo {author} {\bibfnamefont {A.}~\bibnamefont {Denner}}, \
  and\ \bibinfo {author} {\bibfnamefont {S.}~\bibnamefont {Dittmaier}},\ }\href
  {\doibase 10.1103/PhysRevLett.99.161803} {\bibfield  {journal} {\bibinfo
  {journal} {Phys. Rev. Lett.}\ }\textbf {\bibinfo {volume} {99}},\ \bibinfo
  {pages} {161803} (\bibinfo {year} {2007})},\ \Eprint
  {http://arxiv.org/abs/0707.0381} {arXiv:0707.0381 [hep-ph]} \BibitemShut
  {NoStop}%
\bibitem [{\citenamefont {Ciccolini}\ \emph {et~al.}(2008)\citenamefont
  {Ciccolini}, \citenamefont {Denner},\ and\ \citenamefont
  {Dittmaier}}]{Ciccolini:2007ec}%
  \BibitemOpen
  \bibfield  {author} {\bibinfo {author} {\bibfnamefont {M.}~\bibnamefont
  {Ciccolini}}, \bibinfo {author} {\bibfnamefont {A.}~\bibnamefont {Denner}}, \
  and\ \bibinfo {author} {\bibfnamefont {S.}~\bibnamefont {Dittmaier}},\ }\href
  {\doibase 10.1103/PhysRevD.77.013002} {\bibfield  {journal} {\bibinfo
  {journal} {Phys. Rev.}\ }\textbf {\bibinfo {volume} {D77}},\ \bibinfo {pages}
  {013002} (\bibinfo {year} {2008})},\ \Eprint {http://arxiv.org/abs/0710.4749}
  {arXiv:0710.4749 [hep-ph]} \BibitemShut {NoStop}%
\bibitem [{\citenamefont {Figy}\ \emph {et~al.}(2012)\citenamefont {Figy},
  \citenamefont {Palmer},\ and\ \citenamefont {Weiglein}}]{Figy:2010ct}%
  \BibitemOpen
  \bibfield  {author} {\bibinfo {author} {\bibfnamefont {T.}~\bibnamefont
  {Figy}}, \bibinfo {author} {\bibfnamefont {S.}~\bibnamefont {Palmer}}, \ and\
  \bibinfo {author} {\bibfnamefont {G.}~\bibnamefont {Weiglein}},\ }\href
  {\doibase 10.1007/JHEP02(2012)105} {\bibfield  {journal} {\bibinfo  {journal}
  {JHEP}\ }\textbf {\bibinfo {volume} {02}},\ \bibinfo {pages} {105} (\bibinfo
  {year} {2012})},\ \Eprint {http://arxiv.org/abs/1012.4789} {arXiv:1012.4789
  [hep-ph]} \BibitemShut {NoStop}%
\bibitem [{\citenamefont {Bolzoni}\ \emph {et~al.}(2010)\citenamefont
  {Bolzoni}, \citenamefont {Maltoni}, \citenamefont {Moch},\ and\ \citenamefont
  {Zaro}}]{Bolzoni:2010xr}%
  \BibitemOpen
  \bibfield  {author} {\bibinfo {author} {\bibfnamefont {P.}~\bibnamefont
  {Bolzoni}}, \bibinfo {author} {\bibfnamefont {F.}~\bibnamefont {Maltoni}},
  \bibinfo {author} {\bibfnamefont {S.-O.}\ \bibnamefont {Moch}}, \ and\
  \bibinfo {author} {\bibfnamefont {M.}~\bibnamefont {Zaro}},\ }\href {\doibase
  10.1103/PhysRevLett.105.011801} {\bibfield  {journal} {\bibinfo  {journal}
  {Phys. Rev. Lett.}\ }\textbf {\bibinfo {volume} {105}},\ \bibinfo {pages}
  {011801} (\bibinfo {year} {2010})},\ \Eprint {http://arxiv.org/abs/1003.4451}
  {arXiv:1003.4451 [hep-ph]} \BibitemShut {NoStop}%
\bibitem [{\citenamefont {Bolzoni}\ \emph {et~al.}(2012)\citenamefont
  {Bolzoni}, \citenamefont {Maltoni}, \citenamefont {Moch},\ and\ \citenamefont
  {Zaro}}]{Bolzoni:2011cu}%
  \BibitemOpen
  \bibfield  {author} {\bibinfo {author} {\bibfnamefont {P.}~\bibnamefont
  {Bolzoni}}, \bibinfo {author} {\bibfnamefont {F.}~\bibnamefont {Maltoni}},
  \bibinfo {author} {\bibfnamefont {S.-O.}\ \bibnamefont {Moch}}, \ and\
  \bibinfo {author} {\bibfnamefont {M.}~\bibnamefont {Zaro}},\ }\href {\doibase
  10.1103/PhysRevD.85.035002} {\bibfield  {journal} {\bibinfo  {journal} {Phys.
  Rev.}\ }\textbf {\bibinfo {volume} {D85}},\ \bibinfo {pages} {035002}
  (\bibinfo {year} {2012})},\ \Eprint {http://arxiv.org/abs/1109.3717}
  {arXiv:1109.3717 [hep-ph]} \BibitemShut {NoStop}%
\bibitem [{\citenamefont {Dreyer}\ and\ \citenamefont
  {Karlberg}(2016)}]{Dreyer:2016oyx}%
  \BibitemOpen
  \bibfield  {author} {\bibinfo {author} {\bibfnamefont {F.~A.}\ \bibnamefont
  {Dreyer}}\ and\ \bibinfo {author} {\bibfnamefont {A.}~\bibnamefont
  {Karlberg}},\ }\href {\doibase 10.1103/PhysRevLett.117.072001} {\bibfield
  {journal} {\bibinfo  {journal} {Phys. Rev. Lett.}\ }\textbf {\bibinfo
  {volume} {117}},\ \bibinfo {pages} {072001} (\bibinfo {year} {2016})},\
  \Eprint {http://arxiv.org/abs/1606.00840} {arXiv:1606.00840 [hep-ph]}
  \BibitemShut {NoStop}%
\bibitem [{\citenamefont {Kauer}\ \emph {et~al.}(1999)\citenamefont {Kauer},
  \citenamefont {Reina}, \citenamefont {Repond},\ and\ \citenamefont
  {Zeppenfeld}}]{Kauer:1999px}%
  \BibitemOpen
  \bibfield  {author} {\bibinfo {author} {\bibfnamefont {N.}~\bibnamefont
  {Kauer}}, \bibinfo {author} {\bibfnamefont {L.}~\bibnamefont {Reina}},
  \bibinfo {author} {\bibfnamefont {J.}~\bibnamefont {Repond}}, \ and\ \bibinfo
  {author} {\bibfnamefont {D.}~\bibnamefont {Zeppenfeld}},\ }\href {\doibase
  10.1016/S0370-2693(99)00735-2} {\bibfield  {journal} {\bibinfo  {journal}
  {Phys. Lett.}\ }\textbf {\bibinfo {volume} {B460}},\ \bibinfo {pages} {189}
  (\bibinfo {year} {1999})},\ \Eprint {http://arxiv.org/abs/hep-ph/9904500}
  {arXiv:hep-ph/9904500 [hep-ph]} \BibitemShut {NoStop}%
\bibitem [{\citenamefont {Ball}\ \emph {et~al.}(2015)\citenamefont {Ball} \emph
  {et~al.}}]{Ball:2014uwa}%
  \BibitemOpen
  \bibfield  {author} {\bibinfo {author} {\bibfnamefont {R.~D.}\ \bibnamefont
  {Ball}} \emph {et~al.} (\bibinfo {collaboration} {NNPDF}),\ }\href {\doibase
  10.1007/JHEP04(2015)040} {\bibfield  {journal} {\bibinfo  {journal} {JHEP}\
  }\textbf {\bibinfo {volume} {04}},\ \bibinfo {pages} {040} (\bibinfo {year}
  {2015})},\ \Eprint {http://arxiv.org/abs/1410.8849} {arXiv:1410.8849
  [hep-ph]} \BibitemShut {NoStop}%
\bibitem [{\citenamefont {Cacciari}\ \emph {et~al.}(2008)\citenamefont
  {Cacciari}, \citenamefont {Salam},\ and\ \citenamefont
  {Soyez}}]{Cacciari:2008gp}%
  \BibitemOpen
  \bibfield  {author} {\bibinfo {author} {\bibfnamefont {M.}~\bibnamefont
  {Cacciari}}, \bibinfo {author} {\bibfnamefont {G.~P.}\ \bibnamefont {Salam}},
  \ and\ \bibinfo {author} {\bibfnamefont {G.}~\bibnamefont {Soyez}},\ }\href
  {\doibase 10.1088/1126-6708/2008/04/063} {\bibfield  {journal} {\bibinfo
  {journal} {JHEP}\ }\textbf {\bibinfo {volume} {04}},\ \bibinfo {pages} {063}
  (\bibinfo {year} {2008})},\ \Eprint {http://arxiv.org/abs/0802.1189}
  {arXiv:0802.1189 [hep-ph]} \BibitemShut {NoStop}%
\bibitem [{\citenamefont {Figy}\ \emph {et~al.}(2008)\citenamefont {Figy},
  \citenamefont {Hankele},\ and\ \citenamefont {Zeppenfeld}}]{Figy:2007kv}%
  \BibitemOpen
  \bibfield  {author} {\bibinfo {author} {\bibfnamefont {T.}~\bibnamefont
  {Figy}}, \bibinfo {author} {\bibfnamefont {V.}~\bibnamefont {Hankele}}, \
  and\ \bibinfo {author} {\bibfnamefont {D.}~\bibnamefont {Zeppenfeld}},\
  }\href {\doibase 10.1088/1126-6708/2008/02/076} {\bibfield  {journal}
  {\bibinfo  {journal} {JHEP}\ }\textbf {\bibinfo {volume} {02}},\ \bibinfo
  {pages} {076} (\bibinfo {year} {2008})},\ \Eprint
  {http://arxiv.org/abs/0710.5621} {arXiv:0710.5621 [hep-ph]} \BibitemShut
  {NoStop}%
\bibitem [{\citenamefont {Campanario}\ \emph {et~al.}(2013)\citenamefont
  {Campanario}, \citenamefont {Figy}, \citenamefont {Plätzer},\ and\
  \citenamefont {Sjödahl}}]{Campanario:2013fsa}%
  \BibitemOpen
  \bibfield  {author} {\bibinfo {author} {\bibfnamefont {F.}~\bibnamefont
  {Campanario}}, \bibinfo {author} {\bibfnamefont {T.~M.}\ \bibnamefont
  {Figy}}, \bibinfo {author} {\bibfnamefont {S.}~\bibnamefont {Plätzer}}, \
  and\ \bibinfo {author} {\bibfnamefont {M.}~\bibnamefont {Sjödahl}},\ }\href
  {\doibase 10.1103/PhysRevLett.111.211802} {\bibfield  {journal} {\bibinfo
  {journal} {Phys. Rev. Lett.}\ }\textbf {\bibinfo {volume} {111}},\ \bibinfo
  {pages} {211802} (\bibinfo {year} {2013})},\ \Eprint
  {http://arxiv.org/abs/1308.2932} {arXiv:1308.2932 [hep-ph]} \BibitemShut
  {NoStop}%
\bibitem [{\citenamefont {Arnold}\ \emph {et~al.}(2009)\citenamefont {Arnold}
  \emph {et~al.}}]{Arnold:2008rz}%
  \BibitemOpen
  \bibfield  {author} {\bibinfo {author} {\bibfnamefont {K.}~\bibnamefont
  {Arnold}} \emph {et~al.},\ }\href {\doibase 10.1016/j.cpc.2009.03.006}
  {\bibfield  {journal} {\bibinfo  {journal} {Comput. Phys. Commun.}\ }\textbf
  {\bibinfo {volume} {180}},\ \bibinfo {pages} {1661} (\bibinfo {year}
  {2009})},\ \Eprint {http://arxiv.org/abs/0811.4559} {arXiv:0811.4559
  [hep-ph]} \BibitemShut {NoStop}%
\bibitem [{\citenamefont {Baglio}\ \emph {et~al.}(2014)\citenamefont {Baglio}
  \emph {et~al.}}]{Baglio:2014uba}%
  \BibitemOpen
  \bibfield  {author} {\bibinfo {author} {\bibfnamefont {J.}~\bibnamefont
  {Baglio}} \emph {et~al.},\ }\href@noop {} {\  (\bibinfo {year} {2014})},\
  \Eprint {http://arxiv.org/abs/1404.3940} {arXiv:1404.3940 [hep-ph]}
  \BibitemShut {NoStop}%
\bibitem [{\citenamefont {Baglio}\ \emph {et~al.}(2017)\citenamefont {Baglio},
  \citenamefont {Bellm}, \citenamefont {Bozzi}, \citenamefont {Brieg},
  \citenamefont {Campanario}, \citenamefont {Englert}, \citenamefont {Feigl},
  \citenamefont {Frank}, \citenamefont {Figy}, \citenamefont {Geyer},
  \citenamefont {Hackstein}, \citenamefont {Hankele}, \citenamefont {Jäger},
  \citenamefont {Kerner}, \citenamefont {Kubocz}, \citenamefont {Löschner},
  \citenamefont {Ninh}, \citenamefont {Oleari}, \citenamefont {Palmer},
  \citenamefont {Plätzer}, \citenamefont {Rauch}, \citenamefont {Roth},
  \citenamefont {Rzehak}, \citenamefont {Schissler}, \citenamefont
  {Schlimpert}, \citenamefont {Spannowsky}, \citenamefont {Worek},\ and\
  \citenamefont {Zeppenfeld}}]{VBFNLO}%
  \BibitemOpen
  \bibfield  {author} {\bibinfo {author} {\bibfnamefont {J.}~\bibnamefont
  {Baglio}}, \bibinfo {author} {\bibfnamefont {J.}~\bibnamefont {Bellm}},
  \bibinfo {author} {\bibfnamefont {G.}~\bibnamefont {Bozzi}}, \bibinfo
  {author} {\bibfnamefont {M.}~\bibnamefont {Brieg}}, \bibinfo {author}
  {\bibfnamefont {F.}~\bibnamefont {Campanario}}, \bibinfo {author}
  {\bibfnamefont {C.}~\bibnamefont {Englert}}, \bibinfo {author} {\bibfnamefont
  {B.}~\bibnamefont {Feigl}}, \bibinfo {author} {\bibfnamefont
  {J.}~\bibnamefont {Frank}}, \bibinfo {author} {\bibfnamefont
  {T.}~\bibnamefont {Figy}}, \bibinfo {author} {\bibfnamefont {F.}~\bibnamefont
  {Geyer}}, \bibinfo {author} {\bibfnamefont {C.}~\bibnamefont {Hackstein}},
  \bibinfo {author} {\bibfnamefont {V.}~\bibnamefont {Hankele}}, \bibinfo
  {author} {\bibfnamefont {B.}~\bibnamefont {Jäger}}, \bibinfo {author}
  {\bibfnamefont {M.}~\bibnamefont {Kerner}}, \bibinfo {author} {\bibfnamefont
  {M.}~\bibnamefont {Kubocz}}, \bibinfo {author} {\bibfnamefont
  {M.}~\bibnamefont {Löschner}}, \bibinfo {author} {\bibfnamefont {L.~D.}\
  \bibnamefont {Ninh}}, \bibinfo {author} {\bibfnamefont {C.}~\bibnamefont
  {Oleari}}, \bibinfo {author} {\bibfnamefont {S.}~\bibnamefont {Palmer}},
  \bibinfo {author} {\bibfnamefont {S.}~\bibnamefont {Plätzer}}, \bibinfo
  {author} {\bibfnamefont {M.}~\bibnamefont {Rauch}}, \bibinfo {author}
  {\bibfnamefont {R.}~\bibnamefont {Roth}}, \bibinfo {author} {\bibfnamefont
  {H.}~\bibnamefont {Rzehak}}, \bibinfo {author} {\bibfnamefont
  {F.}~\bibnamefont {Schissler}}, \bibinfo {author} {\bibfnamefont
  {O.}~\bibnamefont {Schlimpert}}, \bibinfo {author} {\bibfnamefont
  {M.}~\bibnamefont {Spannowsky}}, \bibinfo {author} {\bibfnamefont
  {M.}~\bibnamefont {Worek}}, \ and\ \bibinfo {author} {\bibfnamefont
  {D.}~\bibnamefont {Zeppenfeld}},\ }\href@noop {} {\enquote {\bibinfo {title}
  {{VBFNLO} - a parton level {Monte Carlo} for processes with electroweak
  bosons},}\ }\bibinfo {howpublished} {\url{https://www.itp.kit.edu/vbfnlo}}
  (\bibinfo {year} {2017})\BibitemShut {NoStop}%
\end{thebibliography}%

\end{document}